\documentclass[journal]{IEEEtran}

\usepackage{amsmath,amsfonts}
\usepackage{algorithmic}
\usepackage{algorithm}
\usepackage{textcomp}
\usepackage{stfloats}
\usepackage{url}
\usepackage{verbatim}
\usepackage{graphicx}
\usepackage{cite}
\usepackage{color}
\usepackage[tight]{subfigure}

\usepackage{acronym}
\newacro{gfm}[GFM]{Grid-Forming}
\newacro{gfl}[GFL]{Grid-Following}
\newacro{ibr}[IBR]{Inverter-Based Resource}
\newacro{pll}[PLL]{Phase-Locked Loop}
\newacro{vsm}[VSM]{Virtual Synchronous Machine}
\newacro{sm}[SM]{Synchronous Machine}
\newacro{cf}[CF]{Complex Frequency}
\newacro{coi}[CoI]{Center of Inertia}
\newacro{rocof}[RoCoF]{Rate of Change of Frequency}
\newacro{dfig}[DFIG]{Doubly-Fed Induction Generator}
\newacro{gvr}[GVR]{Global Voltage Regulation}
\newacro{qss}[QSS]{Quasi-Steady State}
\newacro{pmu}[PMU]{Phasor Measurement Unit}

\begin{document}

\title{On the Local and Global Nature of Frequency and Voltage Dynamics}

\author{Taulant K\"{e}r{\c{c}}i,~\IEEEmembership{Senior Member,~IEEE}, and Federico Milano,~\IEEEmembership{Fellow,~IEEE}
   \thanks{T.~K{\"e}r{\c c}i is with the Irish Transmission System Operator,
     EirGrid, Ballsbridge, D04FW28, Ireland.}
   \thanks{F.~Milano is with School of Electrical and Electronic Engineering,
    University College Dublin, Belfield Campus, D04V1W8, Ireland.
    Corresponding author's e-mail: federico.milano@ucd.ie.}
  
  \thanks{This work was partially supported by Sustainable Energy Authority of Ireland (SEAI) by funding F.~Milano through FRESLIPS project, Grant No.~RDD/00681.}
  \vspace{-5mm}
}

\maketitle

\begin{abstract}
  This paper utilizes the autocorrelation of frequency and voltage measurements to identify, quantify and classify local and global properties of power system dynamics.  
  The analysis is based on measurements with various resolutions (20 ms, 1 s, and 1 min) from several nodes of the Irish All-Island Power System (AIPS).  Simulations based on stochastic differential algebraic equations on an IEEE benchmark system support conclusions drawn from real-world data.
\end{abstract}


\section{Introduction}
\label{sec:intro}

In conventional power systems, it is customary to assume the dynamics of voltages as `local', whereas the dynamic of the frequency is generally assumed to be a `global' trend of the system.  With the integration of inverter-based resources (IBRs) and their fast controllers, local frequency and voltage dynamics may overlap in the time frame of tens of hundreds of milliseconds \cite{8450880}.  However, it is still not clear how this coupling modifies the local vs global properties of frequency and voltage dynamics \cite{9796617}.  This paper addresses this issue through a statistical approach based on the evaluation of the autocorrelation of the frequency and the voltage magnitude at nodes at different locations of the grid.  The objective is to provide a metric able to identify the time scales of local and global dynamics of the measured quantities.  

Recently, the authors proposed new frequency quality metrics based on a time-dependent metric namely the autocorrelation of the frequency (ACF) that allow quantifying the impact of fast dynamics on frequency quality \cite{11578241}.   This work proposes the application of the ACF and autocorrelation of the voltage (ACV) as metrics to fill the above gaps.  Specifically, the ACF and the ACV are applied to a number of diverse buses of a real-world IBR-based system namely the All-Island power system (AIPS) of Ireland.  Dynamic stochastic simulations are then run to support conclusions based on the observation of real-world measurement data.

\section{Autocorrelation Function}
\label{sec:math}

The autocorrelation function measures how much a signal or time series correlates with a delayed version of itself across different time lags ($\tau$).  Instead of comparing two different variables (cross-correlation), it tests how strongly a variable's past values predict its future values.

The autocorrelation of an exponentially decaying signal is also exponentially decaying with the same time constant, and the average autocorrelation of a periodic signal is also periodic.  For this reason, the autocorrelation can be  utilized to reveal repeating cycles hidden inside noisy signals, e.g., seasonality in wind data \cite{JONSDOTTIR2019368}. 

In this work, we utilize the feature of the autocorrelation of deterministic signal that consists in allowing a clear separation and identification of dynamics occurring at different time scales.  To illustrate this property, consider a real-valued signal $x(t)$ composed of a slowly varying baseline $h(t)$ and a damped harmonic transient component $y(t)$:
\begin{equation}
  \label{eq:signal}
  x(t) = h(t) + y(t) = h(t) + a e^{-\alpha t} \cos(\omega_0 t + \phi) u(t) ,
\end{equation}
where $u(t)$ is the unit step function, $\alpha > 0$, $\omega_0 > 0$, and $a \ll h(t) \approx 1$.  Moreover, $h(t)$ varies very slowly relative to the oscillation period and decay time of the transient term.  The values of the parameters $\alpha$ and $\omega_0$ are such that they represent typical damped modes in power systems transients. 

The total autocorrelation $R_{xx}$ expands into four terms:
\begin{equation}
  R_{xx}(\tau) = R_{hh}(\tau) + R_{hy}(\tau) + R_{yh}(\tau) + R_{yy}(\tau) ,
\end{equation}
where $R_{ij}(\tau) = \int_{-\infty}^{\infty} i(t) \, j(t-\tau) \, dt$.
Since $h(t) \approx 1$ is quasi-constant over the decay duration of $x(t)$, its autocorrelation represents a slowly varying baseline, say $R_{hh}(\tau)$.
%

Then, for $\tau \ge 0$, the autocorrelation of the transient term $y(t)$ is given by:
\begin{equation}
  \begin{aligned}
  R_{yy}(\tau) = &\; \frac{a^2 e^{-\alpha \tau}}{4\alpha (\alpha^2 + \omega_0^2)} \big [ (\alpha^2 + \omega_0^2) \cos(\omega_0 \tau) \; + \\
  &\; \alpha^2 \cos(\omega_0 \tau + 2\phi) - 
  \alpha \omega_0 \sin(\omega_0 \tau + 2\phi) \big ] ,         
  \end{aligned}
\end{equation}
which is thus a term that evolves in time as the original signal $y(t)$, that is, it shows damped oscillations with damping $\alpha$ and angular frequency $\omega_0$.  For sake of illustration, let us assume that $\alpha \ll \omega_0$, which is typical of electromechanical oscillations.  This leads to the simplified expression:
\begin{equation}
   R_{yy}(\tau) \approx \frac{a^2}{4\alpha} e^{-\alpha |\tau|} \cos(\omega_0 \tau) .
\end{equation}

The cross-term $R_{hy}(\tau)$ integrates the nearly constant $h(t) \approx 1$ against $y(t)$:
\begin{equation}
R_{hy}(\tau) \approx a \int_{\tau}^{\infty} e^{-\alpha(t-\tau)} \cos\left(\omega_0(t-\tau) + \phi\right) \, dt ,
\end{equation}
or, equivalently:
\begin{equation}
\begin{aligned}
    R_{hy}(\tau) 
    &\approx \frac{a}{\alpha^2 + \omega_0^2} \left( \alpha \cos \phi - \omega_0 \sin \phi \right) ,
\end{aligned}
\end{equation}
which is constant, with $R_{hy} \ll h(t)$ as $a\ll 1$ and, typically $\omega_0 > 1$ or even $\omega_0 \gg 1$ for faster controllers of power electronic converters.  Then, note that $R_{yh}(\tau) = R_{hy}(-\tau)$, hence the condition $R_{yh} \ll 1$ holds as well. 

Combining all non-negligible terms, the total autocorrelation function $R_{xx}$ is:
\begin{equation}
  \label{eq:ac}
  \boxed{R_{xx}(\tau) \approx R_{hh}(\tau) + \frac{a^2}{4\alpha} e^{-\alpha |\tau|} \cos(\omega_0 \tau)}
\end{equation}

The resulting autocorrelation consists of two components:
\begin{itemize}
\item A slowly varying baseline pedestal $R_{hh}(\tau)$ reflecting the low-frequency behavior of $h(t)$.
\item A high-frequency exponentially decaying oscillation $\frac{a^2}{4\alpha} e^{-\alpha |\tau|} \cos(\omega_0 \tau)$ riding on top of the baseline, fully characterizing the transient dynamic mode.
\end{itemize}

On the other hand, the autocorrelation filters completely the effect of pure random noise.  White noise has in fact zero correlation at any non-zero lag, i.e., $R(\tau) = 0, \, \forall \tau \neq 0$.  Thus, if a time series' residual autocorrelation drops to near zero, the remaining variance represents noise.  

The main idea of this work is to utilize the autocorrelation of real-world data to identify dominant modes, either exponentially decaying or periodic, of that time series.  The modes that are local will be specific of a bus, whereas the modes that are common to the system will be visible at different buses.  Moreover, the behavior of the autocorrelation allows defining in which time scales dynamics are local/global.

\section{Real-World Data}
\label{sec:real}

To calculate the ACFs and ACVs, we utilize 3-hour-worth time-series (10,800 s) of voltage and frequency measurements from 5 buses in the  AIPS from 1st of May 2026.  The period of the measurements has been randomly selected among periods of normal operation of the system, that is, no major contingency occurs in the three hours.  The buses of the Irish system are chosen in such a way that they are geographically distant from each other and represent different types of loads, generators, as well as weak and strong areas of the grid.  Three time resolutions are used: 20 ms, 1 s and 1 min.  

The evaluation of the autocorrelation from data does not require to know the analytical expression of the measured signal.  For a continuous, stationary signal $X(t)$ with mean $\mu$ and variance $\sigma^2$, the autocorrelation can be estimated as:
\begin{equation}
    R(\tau) = \frac{E[(X(t) - \mu)(X(t + \tau) - \mu)]}{\sigma^2} \, ,
\end{equation}
where $E(\cdot)$ is the expectation of the signal.

\vspace{-2mm}

\subsection{Autocorrelation of Bus Voltage Frequency (ACF)}

\begin{figure}[htb]
  \begin{center}
    \resizebox{0.80\linewidth}{!}{\includegraphics{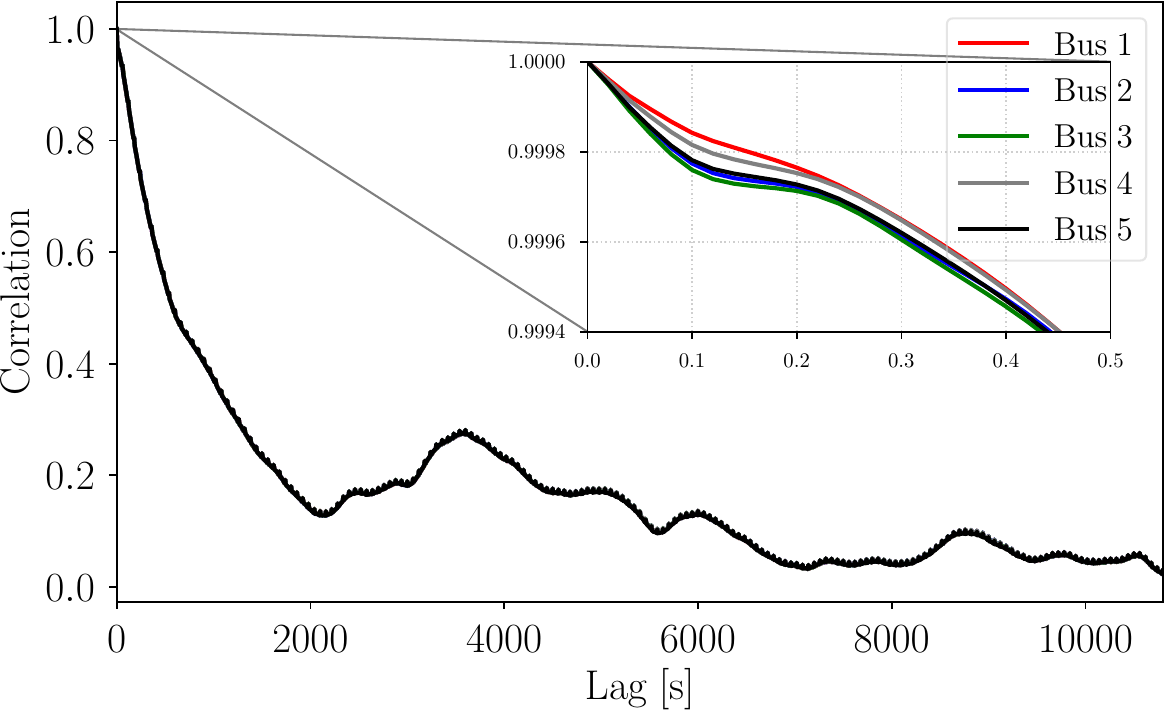}}
    \caption{ACFs of different buses using 20 ms resolution.}
    \label{fig:freq_acfs1}
  \end{center}
  \vspace{-4mm}
\end{figure}

Figure \ref{fig:freq_acfs1} shows the ACFs of the five considered buses using 20 ms resolution.  The goal is to see if there are any differences across different buses and different timescales.  While differences are necessarily small, ACF does not show the same behavior in the five buses in the first 500 ms.  The observed differences are not due to different PMU vendors or types.  As measurements were taken in an interval of time during which no major disturbance occurred in the system, results indicate that frequency has local components during normal operating conditions.  After 550 ms, the ACFs of the five buses substantially coincide.  Recalling \eqref{eq:signal} and \eqref{eq:ac}, one can also conclude that for time frames bigger than 500 ms, the slow component $h(t)$ of the frequencies is substantially the same everywhere in the grid, which is consistent with common experience.   Finally,  the change of slope of the ACF at around 30 minutes is due to market operations in the AIPS. 

\subsection{Autocorrelation of Bus Voltage Magnitude (ACV)}

\begin{figure}[htb]
  \centering
  \includegraphics[width=0.80\linewidth]{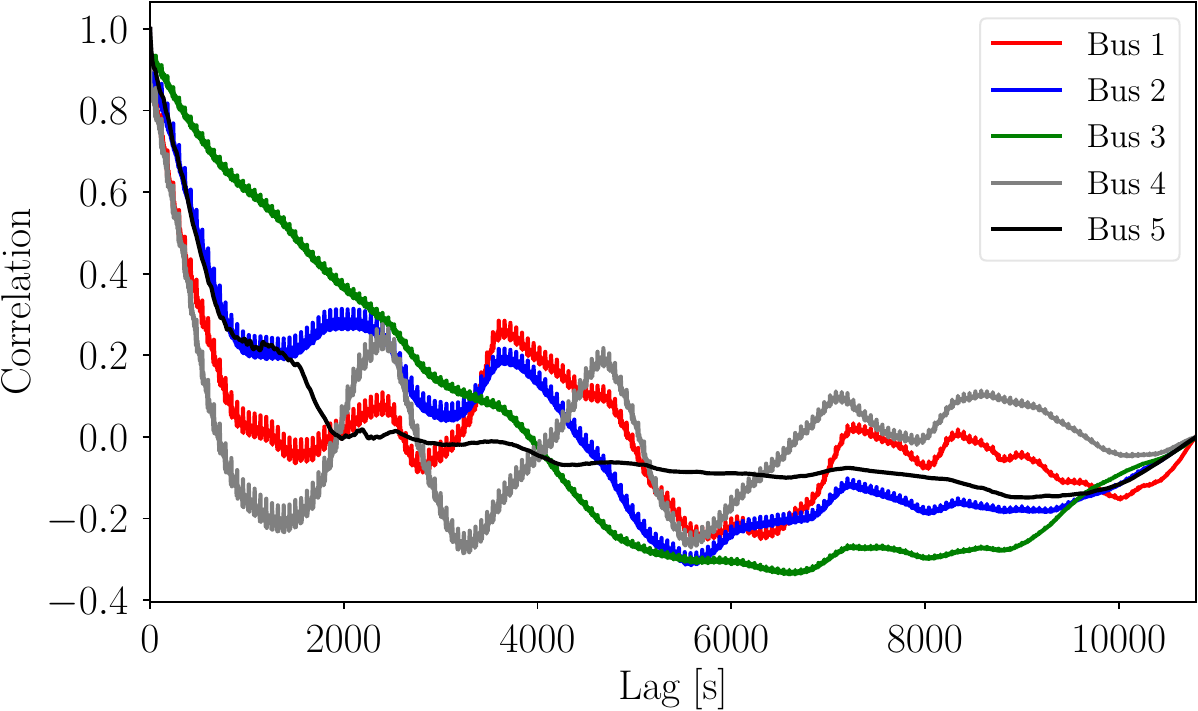}
  \caption{ACVs of different buses using 20 ms resolution.}
  \label{fig:acv}
  \vspace{-2mm}
\end{figure}

\begin{figure}[htb]
  \centering
  \includegraphics[height=3.9cm]{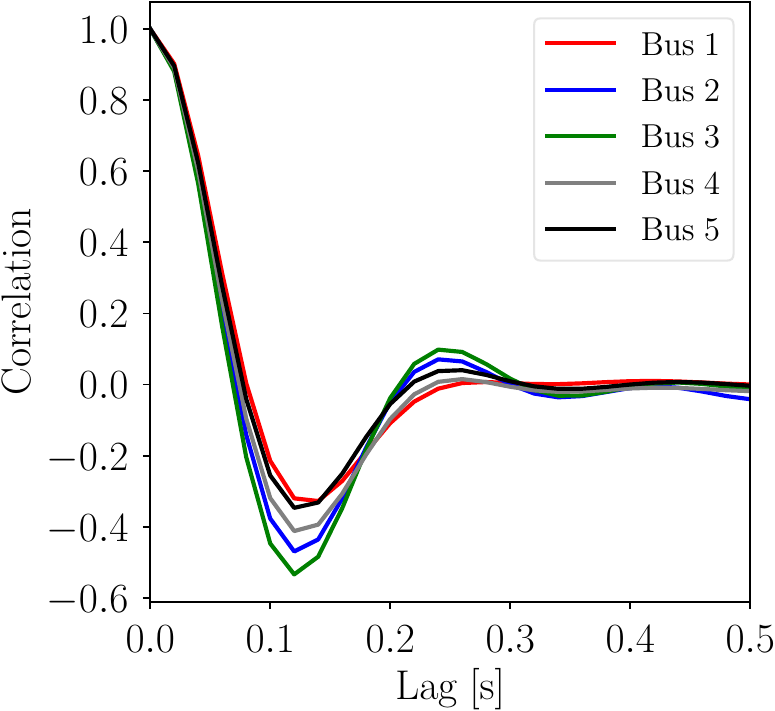}
  \includegraphics[height=3.9cm]{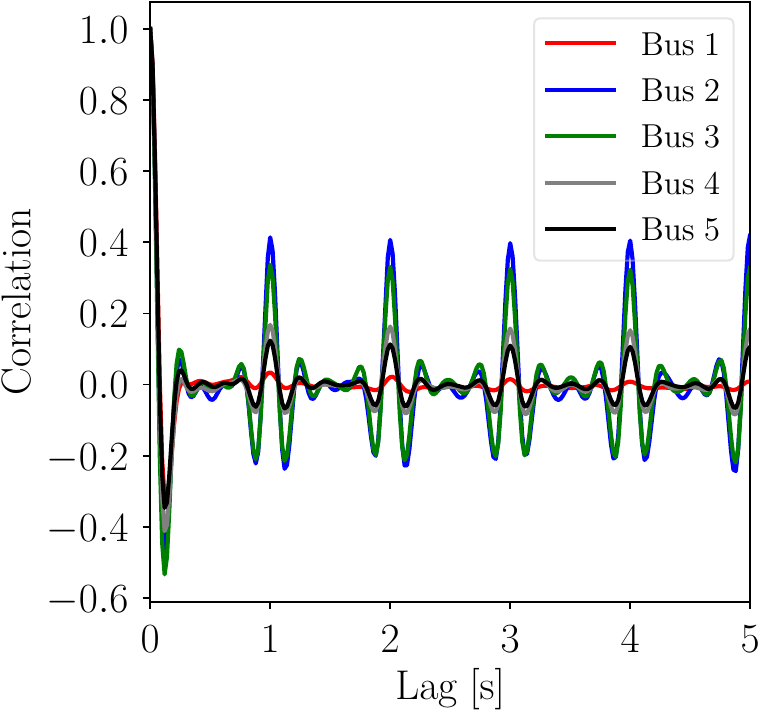}  
  \caption{Autocorrelation of the rate of change of voltage magnitude at different buses using 20 ms resolution.}
  \label{fig:rho}
\end{figure}

Figure \ref{fig:acv} shows the ACVs of the five considered buses using 20 ms resolution.  The long term components of the ACVs are different depending on the location of the bus.  This confirms that voltage magnitude variations are essentially driven by local dynamics.  However, Fig.~\ref{fig:rho} shows that the autocorrelation of the rates of change of bus voltage magnitudes is similar to that of frequency in the short-term time scales.  That is, there are fast local modes in the time scale of 100 ms whereas in the time scale of 1 s voltages show same behavior regardless of their geographical position.  Note also the 1 Hz cycles due to the data centers connected to the AIPS.


\begin{figure}[htb]
  \begin{center}
    \resizebox{0.80\linewidth}{!}{\includegraphics{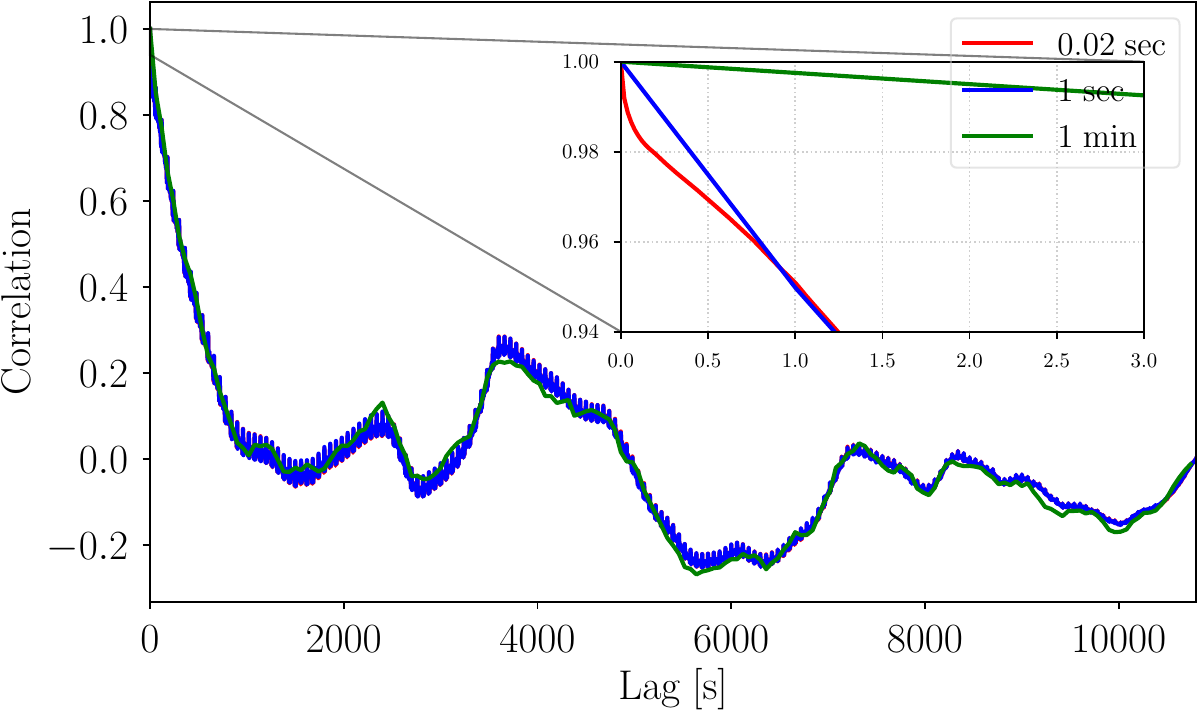}}
    \caption{ACV of bus 1 using 20 ms, 1 s and 1 min resolutions.}
    \label{fig:volt_acv_diff}
  \end{center}
  \vspace{-3mm}
\end{figure}

Figure \ref{fig:volt_acv_diff} shows the ACV of bus 1 using three different data resolutions.  At higher resolutions, the ACV captures faster dynamics, as expected.  All ACVs of a given bus eventually overlap regardless of the resolution of the measurements.  The sampling time, thus, can be tuned based on the time frame of interest.  Finally, note the ripple visible on the measurements with 0.02 s and 1 s resolution.  This ripple is again due to data center duty cycles with 60 s period.

\section{Case Study}
\label{sec:case}

To confirm the conclusions drawn based on real-world data, we use the IEEE 9-bus system and run 3-hour dynamic stochastic simulations using Dome \cite{6672387}.  The synchronous generator connected at bus 3 is replaced with a wind power plant with the capability of providing both primary frequency and voltage control.  Three scenarios of the wind power plant voltage control are simulated: (i) no voltage control; (ii) slow voltage control; and (iii) fast voltage control.  The third scenario is equivalent to the operating conditions of the AIPS in the period during which the measurements discussed in the previous section were taken.  Primary frequency control and noise are the same in the three cases and no automatic secondary control is included as this is not present in the AIPS.  

\begin{figure}[htb]
  \centering
  \includegraphics[width=0.8\linewidth]{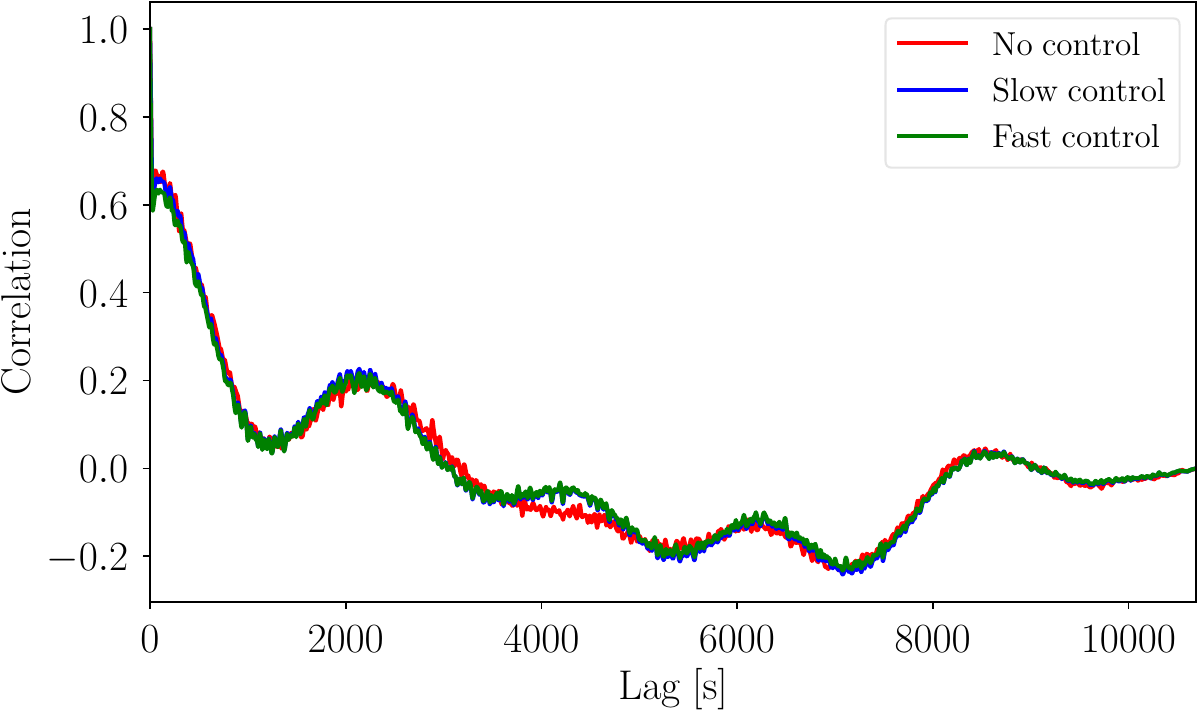}
  \caption{ACVs of bus 3 for different simulated scenarios.}
  \label{fig:sim_acv}
\end{figure}

\begin{figure}[htb]
  \centering
  \includegraphics[width=0.495\linewidth]{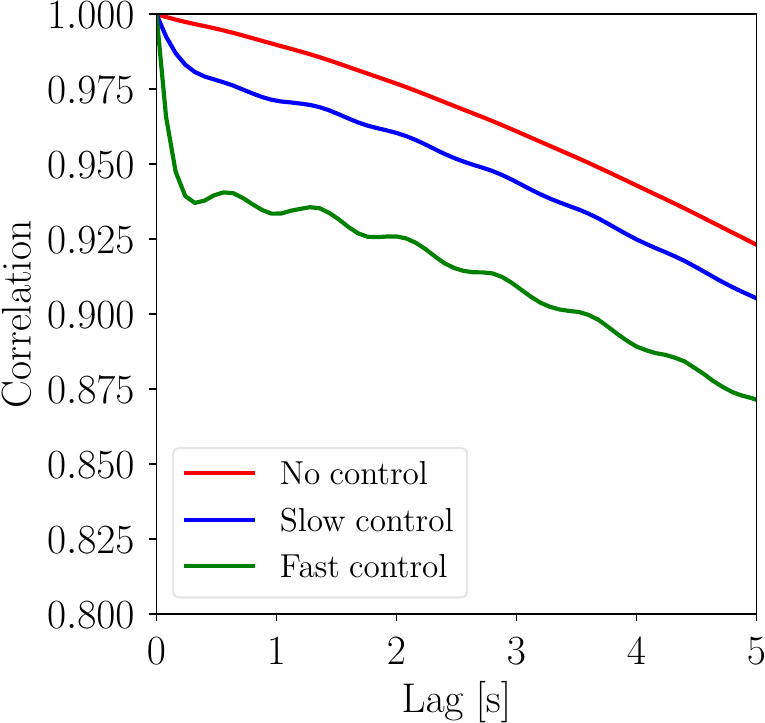}
  \includegraphics[width=0.47\linewidth]{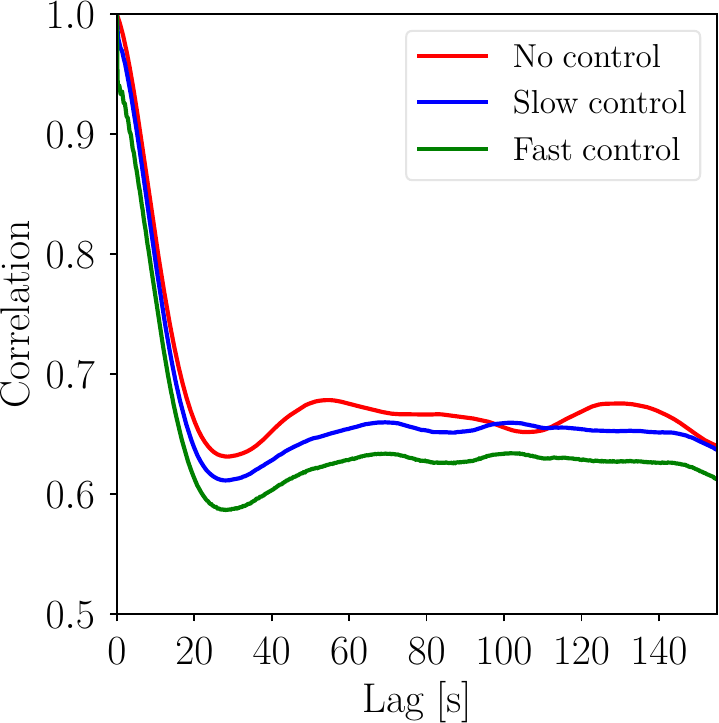}
  \caption{Zooms of ACVs of bus 3 for different time frames and scenarios.}
  \label{fig:sim_acv_zoom}
\end{figure}

Figure \ref{fig:sim_acv} shows a good agreement, in the long term, between simulations and measurements.  Figure \ref{fig:sim_acv_zoom} focuses on short term and medium term dynamics of the voltage and shows that the scenario of fast voltage control leads to initial faster decaying of ACV.  This is due to the fact that the fast control is able to quickly compensate voltage variations and, thus, the voltage changes quicker compared to the cases of slow control and no control.  These results are consistent with real-world observations made in the previous section.

\section{Conclusions}
\label{sec:conclu}

The letter proposes the autocorrelation function as a metric to quantify local and global dynamics of frequency and voltage.  Utilizing operational data from the real-world IBR-based AIPS, the letter demonstrates that:

\begin{itemize}
    \item  Frequency and voltage are both driven by local dynamics in the very short term (100 ms time scale).  This behavior is shown in normal operating conditions and is thus not due to dynamics triggered by large disturbances.  
    \item In the long term (minutes to hours) frequency is mostly a global quantity, whereas voltage magnitude is a local one.   This conclusion is in agreement with common experience.  This conclusion applies for relatively slow dynamics, which are characterized by time scales above 1 s. 
    \item Voltage dynamics have a specific set of dynamics around the 1 s time scale that appear to be common to the whole system.  In this time scale, thus, both frequency and voltage are driven by global dynamics.  This observation is a novel insight of this work.  
\end{itemize}

Simulation results also show that faster voltage control during normal conditions leads to a faster initial decay of the ACV.  The same conclusion can be drawn for fast frequency control.

Future work will focus on the application of the autocorrelation metric to other variables to enable a consistent comparison of the relevant dynamics of different real-world power systems.

\bibliographystyle{IEEEtran}
\bibliography{refs}

@ARTICLE{11578241,
  author={Kërçi, Taulant and others},
  journal={IEEE Trans. on Power Systems}, 
  title={Frequency Quality Metrics based on Second-Order Derivative and Autocorrelation}, 
  year={2026},
  volume={},
  number={},
  pages={1-4},
  doi={10.1109/TPWRS.2026.3707412}}

@article{JONSDOTTIR2019368,
title = {Data-based continuous wind speed models with arbitrary probability distribution and autocorrelation},
journal = {Renewable Energy},
volume = {143},
pages = {368-376},
year = {2019},
issn = {0960-1481},
doi = {https://doi.org/10.1016/j.renene.2019.04.158},
author = {Guðrún Margrét Jónsdóttir and others},
}

@INPROCEEDINGS{6672387,
  author={Milano, Federico},
  booktitle={IEEE PES General Meeting}, 
  title={A {P}ython-based software tool for power system analysis}, 
  year={2013},
  volume={},
  number={},
  pages={1-5},
  doi={10.1109/PESMG.2013.6672387}}

@INPROCEEDINGS{8450880,
  author={Milano, Federico and others},
  booktitle={2018 Power Systems Computation Conference (PSCC)}, 
  title={Foundations and Challenges of Low-Inertia Systems}, 
  year={2018},
  volume={},
  number={},
  pages={1-25},
  doi={10.23919/PSCC.2018.8450880}}

@ARTICLE{9796617,
  author={Gu, Yunjie and others},
  journal={Proc. of the IEEE}, 
  title={Power System Stability With a High Penetration of Inverter-Based Resources}, 
  year={2023},
  volume={111},
  number={7},
  pages={832-853},
  doi={10.1109/JPROC.2022.3179826}}

\end{document}